\documentclass[12pt]{article}

\newcommand{\sect}[1]{\setcounter{equation}{0}\section{#1}\indent}

\newcommand\npb[3]    {
		{{\it Nucl.\ Phys.\ }{\bf B #1} (#2) #3}}
\newcommand\plb[3]   {
		{{\it Phys.\ Lett.\ }{\bf B #1} (#2) #3}}

\newcommand\prl[3]   {
		{{\it Phys.\ Rev.\ Lett.\ }{\bf #1} (#2) #3}}

\newcommand\jhep[3]  {
		{{\it J. High Energy Phys.\ }{\bf #1} (#2) #3}}	
\newcommand{\hepth}[1]{
        {\tt hep-th/#1}}
\usepackage{amsbsy,amssymb,latexsym}

\newcommand{\nn}{\nonumber}
\newcommand{\bref}[1]{(\ref{#1})}
\newcommand{\cint}{{\oint\hspace{-4.4mm}-}}
\newcommand{\cI}{{\cal I}}
\newcommand{\Psib}{\bar\Psi}
\newcommand{\etat}{\tilde\eta}
\newcommand{\Omegab}{\bar\Omega}

\begin{document}

\thispagestyle{empty}
\setcounter{page}{0}
\begin{flushright}
KEK-TH-796\\
hep-th/0112135\\
December 2001
\end{flushright}

\vspace{5mm}

\begin{center}
{\Large \bf{\sc Pregeometrical Formulation\\
\medskip
of\\
\medskip
Berkovits' Open RNS\\
\smallskip
Superstring Field Theories
}}
\end{center}

\vspace{10mm}

\begin{center}
{\large Makoto Sakaguchi}
\bigskip

\textit{Theory Division, Institute of Particle and Nuclear Studies\\
High Energy Accelerator Research Organization (KEK)\\
1-1 Oho, Tsukuba, Ibaraki, 305-0801, JAPAN}\\
\medskip

\texttt{Makoto.Sakaguchi@kek.jp}
\end{center}
\vspace{20mm}

\begin{center}
{\large \textbf{Abstract}}
\end{center}

We propose a pregeometrical formulation of Berkovits' open Ramond-Neveu-Schwarz(RNS)
superstring field theories.
We show that Berkovits' open RNS superstring field theories arise by expanding around
particular solutions of the classical equations of motion for this theory.
Our action contains pure ghost operators only and so is formally background independent.
\newpage
\sect{Introduction}
Tachyon condensation has attracted great interests
after the proposal given by Sen \cite{S;Universality}. 
It is known that the number of space-time supersymmetry
and the space-time geometry changes during tachyon condensation.
A classical analysis of this process
was given in \cite{HS} assuming Sen's conjecture.
In order to investigate this process without this assumption,
one needs an off-shell formulation of superstring theory
of the Gliozzi-Scherk-Olive (GSO) (--) sector as well as the GSO(+) sector,
because the vacuum structure changes during tachyon condensation.
If one would like to extract some information
on space-time supersymmetry,
the Ramond sector must be included in the study.
For this purpose, GSO unprojected superstring field theories
of a Neveu-Schwarz (NS) string field
and a Ramond (R) string field is expected to be powerful.

\medskip

Right after the construction of open bosonic string field theory
(SFT) \cite{W;Interacting}
of a fermionic string field $V$ of $-1$ ghost number,
Witten extended it to open superstring field theory(super-SFT)
 \cite{W;Non-commutative}
of an NS string field $V_{NS}$ of $-1$ picture and a R string field $V_{R}$
 of $-\frac{1}{2}$ picture.
The NS interaction term of the theory must contain the picture changing operator $Z$ 
of $+1$ picture because the non-vanishing small Hilbert space norm
$\left\langle c\partial c\partial^2ce^{-2\phi}\right\rangle=1$
carries $-2$ picture. 
It was shown \cite{W;Scattering} that the collision of two $Z$'s
causes the contact term divergences which prevent the action
from being gauge-invariant.
Berkovits gave a solution \cite{B;Super-Poincare Invariant,B;Review,B;New}
to this contact term problem,
which was to use a bosonic NS string field $\Phi$ of zero picture
and to construct the open NS super-SFT action
to be a Wess-Zumino-Witten like form in the large Hilbert space.
It is not possible to construct an action
for a single R string field $\Psi$
of zero ghost number and $\frac{1}{2}$ picture.
The kinetic term of such a R string field
must contain the inverse picture changing operator $Y$
of $-1$ picture in the large Hilbert space.
The inverse picture changing operator $Y$ causes inconsistencies
due to the non-trivial kernel.
Berkovits provided \cite{B;Super-Poincare Invariant} the N=1 d=4 super-Poincare invariant
open super-SFT action including R string fields as well as an NS string field,
where he used an alternative method which is to split a R string field $\Psi$
into two string fields,
$\Psi$ and $\bar\Psi$, with $\frac{1}{2}$ and $-\frac{1}{2}$ picture respectively.
However it was not clear how to extend this construction to the other dimensions.
Recently in \cite{B;Ramond}, Berkovits beautifully extended  this construction
to the other dimensions
and provided actions of open RNS super-SFT's.
In this paper, we work with this theory.

\medskip

It was shown \cite{HLRS;Purely}
that Witten's bosonic SFT can be obtained from a pure cubic SFT
which does not contain a Becchi-Rouet-Stora-Tyupin (BRST) operator
and so is formally background independent.
Thus pure cubic SFT can be viewed as a fundamental
formulation of Witten's bosonic SFT.
Recently, Berkovits' NS super-SFT
was shown to be formulated pregeometrically by Kluso\v{n} in \cite{K;Proposal}.
The pregeometrical action contains a pure ghost operator $Q_0$
which is related to the BRST operator $Q$ by a similarity transformation
and has non-trivial cohomology.
So this theory describes physical open string excitations.

\medskip

In this paper, we address a question
whether Berkovits' RNS super-SFT's can be formulated pregeometrically.
The pregeometrical super-SFT will contain a pure ghost operator $Q_0$
as in \cite{K;Proposal}.
Berkovits' d=8,4 Lorentz covariant open RNS super-SFT's
contain four operators while
we have only two background independent operators $Q_0$ and the zero mode of $\eta$ .
So we must `create' two operators
as was done in the pure cubic SFT where 
the BRST operator was absent at first and derived
by expanding around solutions of the equations of motion.
This implies that the pregeometrical action will take a different form
from the actions of Berkovits' d=8,4 Lorentz covariant open RNS super-SFT's
while the Berkovits' NS super-SFT action and the pregeometrical action
had the same form \cite{K;Proposal}.

\medskip

The plan of the present paper is as follows.
In the section 2, some relevant aspects of
Berkovits' open RNS super-SFT's are reviewed.
In the section 3, after reviewing the pregeometrical formulation of
Witten's open bosonic SFT, we propose a pregeometrical action for
manifest d=8,4 Lorentz covariant open RNS super-SFT's reviewed in the section 2.
The last section is devoted to a summary and conclusions.

\sect{Review of Berkovits' Open RNS Superstring Field Theories}
In this section we review some relevant aspects of open RNS super-SFT's
constructed by Berkovits in \cite{B;Ramond}.

The RNS superstring theory is constructed from a combined system
of the c=15 superconformal `matter' system and the c=--15 superconformal ghost 
system of a set of fermionic ghosts $(b,c)$
and a set of bosonic ghosts $(\beta,\gamma)$.
It is convenient to feminize
the bosonic ghosts
$(\beta,\gamma)$ as
$\beta=\partial\xi e^{-\phi}$ and $\gamma=\eta e^{\phi}$.
The physical states are defined by non-trivial elements of the BRST cohomology
in the small Hilbert space.
The BRST operator $Q$ is defined 
\cite{ABC;A Note} by\footnote{We use a symbol
$\oint\hspace{-3mm}-$
as $\oint\frac{dz}{2\pi i}$ throughout this paper.}
\begin{eqnarray}
Q=\cint j,~~
j=e^{-S}(-be^{2\phi}\eta\partial\eta)e^{S},~~
S=\cint(cG_me^{-\phi}\xi
+\frac{1}{4}\partial
e^{-2\phi}\xi\partial\xi c\partial c).
\label{Q}
\end{eqnarray}
This BRST operator is equivalent to the ordinary BRST operator
but the BRST current $j$
is different by a total derivative term
and the OPE of two BRST currents is non-singular, $j(z)j(w)\sim 0$.
\medskip

As was explained in the introduction,
we use three string fields $[\Phi,\Psi,\Psib]$
of $[0,\frac{1}{2},-\frac{1}{2}]$ picture and zero ghost
number.
\footnote{
The picture number operator $P$ and the ghost number operator $J_g$
are defined by $P=\oint\hspace{-3.1mm}-(\xi\eta-\partial\phi)$ and
$J_g=\oint\hspace{-3.1mm}-(\eta\xi+cb)$.
}
Berkovits proposed actions for open RNS super-SFT's
supposing that one can define a conserved C-charge as
\begin{eqnarray}
  \begin{array}{c|ccccccc}
       &\Phi&\Psi&\Psib&(b,c)&(\beta,\gamma)&(\xi, \eta)&e^{n\phi}\\\hline
 \mbox{C-charge}&0&-\frac{1}{3}&\frac{1}{3}&(0,0)&(0,0)&(1,-1)&n\\
  \end{array}
~.
\label{C-charge}
\end{eqnarray}
The non-vanishing large Hilbert space norm
$\left\langle\xi c\partial c\partial^2ce^{-2\phi}\right\rangle=1$
carries $-1$ C-charge.
Assuming that $Q$ does not carry C-charge, the simplest action was constructed as
\begin{eqnarray}
S&=&\bigg\langle
\frac{1}{2}(g^{-1}\tilde\eta g)(g^{-1}Qg)
+\frac{1}{2}\int^1_0dt(\hat g^{-1}\partial\hat g)
  \{\hat g^{-1}\tilde\eta\hat g,\hat g^{-1}Q\hat g\}\nonumber\\
&&~~~+g^{-1}(Q\bar\Psi)g(\tilde\eta\Psi)
-\frac{1}{3}\bar\Psi(Q\bar\Psi)^2
+\frac{1}{3}\Psi(\tilde\eta\Psi)^2
\bigg\rangle,
\label{S;SFT1}\\
&&\textrm{with}~~g=e^\Phi,~~\hat g=e^{\hat\Phi(t)},~~\hat\Phi(0)=0,~~\hat\Phi(1)=\Phi,
\nn
\end{eqnarray}
where $\etat$ stands for the  zero-mode of $\eta$ and
we defined $\{~,~]$ by $\{X,Y]=X\star Y-(-1)^{X\cdot Y}Y\star X$
with $(-1)^{X\cdot Y}=-1$ only for both $X$ and $Y$ are Grassmann odd.
The nonlinear equations of motion
obtained varying this action can be realized as $[-\frac{2}{3},-1,-\frac{4}{3}]$ C-charge
part of
the equation
\begin{eqnarray}
&&(G+A)^2=0,
\label{eom}\\
&&G=G_0+G_{-1},~~~
G_0=Q,~~
G_{-1}=\tilde\eta,\nn\\
&&A=A_0+A_{-\frac{1}{3}}+A_{-\frac{2}{3}},~~~
A_0=g^{-1}(Qg),~~
A_{-\frac{1}{3}}=g^{-1}(Q\bar\Psi)g,~~
A_{-\frac{2}{3}}=\tilde\eta\Psi,~~\nonumber
\end{eqnarray}
where $G_n$ and $A_n$ carry $n$ C-charge
and the rest part of $[0,-\frac{1}{3},-\frac{5}{3}]$ C-charge
is automatically satisfied.
We can define a `chiral' (`anti-chiral') field
$\Omega$ ($\bar\Omega$) by $\etat\Omega=0$ ($Q\Omegab=0$).
Since the cohomologies of $Q$ and $\etat$ are trivial in the large Hilbert space,
$\Omega$ ($\Omegab$) can be written as $\etat\Psi$ ($Q\Psib$) for some $\Psi$
($\Psib$).
Therefore one can treat $\Omega$ and $\Omegab$ as fundamental string fields. 
In terms of $\Omega$ and $\Omegab$, the second line of the action \bref{S;SFT1}
 can be expressed 
as,
$
\left\langle g^{-1}\Omegab g\Omega\right\rangle
-\frac{1}{3}\left\langle\Omegab^3\right\rangle_{\bar F}
+\frac{1}{3}\left\langle\Omega^3\right\rangle_{F}
$,
where we defined
the small Hilbert space norms
$\left\langle \etat(\xi e^{-2\phi}c\partial c\partial^2c)\right\rangle_F=1$
and
$\left\langle Q(\xi e^{-2\phi}c\partial c\partial^2c)\right\rangle_{\bar F}=1$.
\medskip

The C-charge assignment \bref{C-charge}
implies that the world-sheet `matter' fields will carry non-zero C-charge.
This causes the BRST operator $Q$ to carry C-charge in general
because the `matter' superconformal generators $T_m$ and $G_m$
appear in $Q$.
There are two ways to respect the C-charge assignment \bref{C-charge}.
First, we generalize $Q$ to carry C-charge below,
which enables us to construct
manifestly d=8,4 Lorentz covariant open RNS super-SFT's.
Secondly, as will be done in the next section,
we use the action \bref{S;SFT1} with 
the pure ghost operator
$Q_0$,
instead of $Q$, as a pregeometrical formulation.

\medskip

Now we generalize $G=Q+\etat$ to
$G=G_0+G_{-\frac{1}{3}}+G_{-\frac{2}{3}}+G_{-1}$,
which implies that $Q$ can carry non-zero C-charge.
One can show that $
(G+A)^2=0
$
implies consistent equations of motion when
\begin{eqnarray}
&&
A=A_0+A_{-\frac{1}{3}}+A_{-\frac{2}{3}},\label{A}\\
&&A_0=g^{-1}(G_0g),~~
A_{-\frac{1}{3}}=g^{-1}(G_{-\frac{1}{3}}g)+g^{-1}\bar\Omega g,~~
A_{-\frac{2}{3}}=\Omega,\nn
\end{eqnarray}
where we defined a chiral (anti-chiral) string fields $\Omega$ $(\Omegab)$
by $G_{-1}\Omega=0$ $(G_0\Omegab=0)$ which implies that
$\Omega=G_{-1}\Psi$ $(\bar\Omega=G_0\bar\Psi)$ for some $\Psi$ ($\Psib$),
assuming the triviality of the cohomologies of $G_0$ and $G_{-1}$.
The equations of motion can be obtained varying the action
\begin{eqnarray}
S&=&\bigg\langle
\frac{1}{2}(g^{-1}G_{-1} g)(g^{-1}G_0g)
+\frac{1}{2}(g^{-1}G_{-\frac{2}{3}} g)(g^{-1}G_{-\frac{1}{3}}g)\nonumber\\&&
~~~+\frac{1}{2}\int^1_0dt(\hat g^{-1}\partial\hat g)
  \biggl( \{\hat g^{-1}G_{-1}\hat g,\hat g^{-1}G_0\hat g\}
        +\{\hat g^{-1}G_{-\frac{2}{3}}\hat g,\hat g^{-1}G_{-\frac{1}{3}}\hat g\}
  \biggr)
\nonumber\\&&
~~~+g^{-1}\bar\Omega g\Omega
+\bar\Omega g(G_{-\frac{2}{3}}g^{-1})
-\Omega g^{-1}(G_{-\frac{1}{3}}g)\bigg\rangle
\nonumber\\&&
-\bigg\langle
\frac{1}{2}\bar\Omega G_{-\frac{1}{3}}\bar\Omega
+\frac{1}{3}(\bar\Omega)^3
\bigg\rangle_{\bar F}
+\bigg\langle
\frac{1}{2}\Omega G_{-\frac{2}{3}}\Omega
+\frac{1}{3}(\Omega)^3
\bigg\rangle_{F}
\label{S;SFT2}
\end{eqnarray}
where $\left\langle~\right\rangle_{\bar F}$ and $\left\langle~\right\rangle_{F}$
are defined using the small Hilbert space norms\\
$\left\langle G_{-1}(\xi e^{-2\phi}c\partial c\partial^2c)\right\rangle_F=1$
and $\left\langle G_{0}(\xi e^{-2\phi}c\partial c\partial^2c)\right\rangle_{\bar F}=1$.

\medskip

Different choice of the definition of C-charge
leads to different background geometry.
Let us define C-charge as
\begin{eqnarray}
C=P+\frac{1}{3}\cint j_N,~~~
j_N=
\left\{
  \begin{array}{ll}
   \displaystyle\psi^0\psi^9,    &~~\textrm{for d=8},  \\
  \displaystyle i(\psi^4\psi^5+\psi^6\psi^7+\psi^8\psi^9), &~~\textrm{for d=4},    \\
  \end{array}
\right.
\end{eqnarray}
where $P$ is the picture, 
so that world-sheet `matter' fields
$\psi^\pm=\frac{1}{\sqrt{2}}(\psi^0\pm\psi^9)$ for d=8
and $\psi^{\pm j}=\frac{1}{\sqrt{2}}(\psi^{2j+2}\pm i\psi^{2j+3}),~j=1,2,3,$
for d=4
carry $\pm\frac{1}{3}$ C-charge.\footnote{
Similarly we define $x^{\pm}=\frac{1}{\sqrt{2}}(x^0\pm x^9)$ for d=8
and $x^{\pm j}=\frac{1}{\sqrt{2}}(x^{2j+2}\pm ix^{2j+3})$, $j=1,2,3$, for d=4.
}
Since $j_N$ for d=4 is defined to be anti-hermitian,
the string fields and the operators must satisfy
the hermiticity condition,
$
\Phi^\dagger=-\Phi,~
\Psi^\dagger=\Psib,~
G_0^\dagger=G_{-1},~
G_{-\frac{1}{3}}^\dagger=~G_{\frac{2}{3}},
$
while, for d=8, they are independent each other.
For d=4, $j_N$ can be generalized to $j_N=\partial H$
which is the U(1) current on a Calabi-Yau manifold.

The operator $G$, which carries $[0,-\frac{1}{3},-\frac{2}{3},-1]$ C-charge,
was defined by a similarity transformation of $Q+\etat$ as
\begin{eqnarray}
G&=&e^R(Q+\etat)e^{-R}~~~~~~\textrm{with}~~
R=
\cint c\xi e^{-\phi}\psi^{+j}\partial x^{-j}
,
\nn\\
&=&
e^{-U}Q_0e^U
+\cint\eta e^\phi\psi^{-j}\partial x^{+j}
+\cint ce^{-\phi}\psi^{+j}\partial x^{-j}
+\etat,
\label{G;2}
\end{eqnarray}
where
$(\psi^{\pm j},x^{\pm j})$ stands for $(\psi^\pm,x^\pm)$
for d=8,
and $U$ is
\begin{eqnarray}
U=
\cint\left(
  c\xi e^{-\phi}
  \psi^p\partial x_p
  +\frac{1}{2}(\partial \phi+
  j_N 
  )c\partial c\xi\partial\xi e^{-2\phi}
\right)
\label{U}
\end{eqnarray}
where $p$ runs from $0$ to $3$ for d=4 and from $1$ to $8$ for d=8.
One can show that the operators $G_{0}$ and $G_{-1}$ are nilpotent and have trivial cohomologies
in the large Hilbert space.
The action \bref{S;SFT2} with $G_n$ obtained above provides
manifest d=8,4 Lorentz covariant open RNS super-SFT's.\footnote{
One can define similarly a d=0 open RNS super-SFT.
}

\medskip

In order to relate to
the manifestly N=1 d=4 super-Poincare covariant open super-SFT action,
Berkovits performed a similarity transformation further
as
$\hat G\equiv
e^{\frac{1}{2}U}Ge^{-\frac{1}{2}U}$
and showed that $\hat G$ can be rewritten
in an N=1 d=4 super-Poincare covariant notation
using Green-Schwarz-like variables
\cite{B;Covariant}.

\sect{Pregeometrical Formulation}
In this section, we will show that Berkovits' RNS super-SFT
reviewed in the previous section
can be formulated pregeometrically.
To do this we begin with reviewing the simplest example
of such a pregeometrical formulation for Witten's bosonic SFT.

\subsection{Pure Cubic String Field Theory}
In \cite{HLRS;Purely},
a pregeometrical SFT action for Witten's open bosonic SFT was proposed as
a pure cubic form
\begin{eqnarray}
S=\left\langle\frac{1}{3} V\star V\star V\right\rangle.
\end{eqnarray}
The equation of motion is
$V\star V=0.$
Expanding around a classical solution $V=V_0+v$,
we obtain an action of the fluctuation $v$
\begin{eqnarray}
S=\left\langle
\frac{1}{2}v\star D_{V_0}v
+\frac{1}{3}v\star v\star v
\right\rangle
\end{eqnarray}
where $D_{V_0}$ is defined by $D_{V_0}X=V_0\star X-(-1)^XX\star V_0$
and is a derivation.
If we can find a classical solution $V_0$ such that $D_{V_0}=Q$,
where $Q$ is the BRST operator of bosonic string theory,
then the action turns out to be the Witten's open bosonic SFT action.
In order to achieve this,
we introduce the half-string formulation \cite{W;Non-commutative}
\begin{eqnarray}
&&{\cal I}\star X=X\star{\cal I}=X,~~~\forall X,\label{half-string;r1}\\
&&Q^R{\cal I}=-Q^L{\cal I},\label{half-string;r2}\\
&&Q^RX\star Y=-(-1)^{X}X\star Q^LY~~~\forall X,Y,
\label{half-string;r3}
\end{eqnarray}
where $Q^{L,R}$ is the BRST operator
integrated over the left/right half of the string
($Q=Q^L+Q^R$) and $\cal I$ is the identity operator of the algebra.
It was shown that 
\begin{eqnarray}
V_0=Q^L{\cal I}
\end{eqnarray}
obeys the equation of motion and satisfies,
$D_{Q^L\cI}X=QX$, for any string field $X$,
using \bref{half-string;r1}-\bref{half-string;r3}
and the fact
\begin{eqnarray}
\{Q,j\}=0,~~~~
Q=\cint j,~~
j=cT_m+bc\partial c+\frac{3}{2}\partial^2c.
\end{eqnarray}
As a result, Witten's SFT has arisen
through a classical solution of pure cubic SFT.

\subsection{Pregeometrical Formulation of
Berkovits' Open RNS Superstring Filed Theories}

In \cite{K;Proposal}, Berkovits' open NS super-SFT was formulated pregeometrically,
where the pregeometrical action includes two operators, $Q_0$ and $\etat$,
and has the same form as the original open NS super-SFT action.
In this section, we will provide a pregeometrical formulation of
manifest d=8,4 Lorentz covariant RNS super-SFT's reviewed in the section 2.

\medskip

As was mentioned in the section 2,
the C-charge assignment \bref{C-charge}
make the world-sheet `matter' fields to carry non-zero C-charge.
This causes the BRST charge $Q$ to carry C-charge in general.
Thus we use the action \bref{S;SFT1} with the pure ghost operator,
$Q_0=-\oint\hspace{-3.8mm}- be^{2\phi}\eta\partial\eta$,
instead of $Q$.
This action is formally background independent
and provides
a pregeometrical formulation
of open RNS super-SFT as will be shown bellow.
\medskip

We propose an action of pregeometrical open RNS super-SFT
\begin{eqnarray}
S&=&\left\langle
  \frac{1}{2}(g^{-1}\tilde\eta g)(g^{-1}Q_0g)
  +\frac{1}{2}\int_0^1dt(\hat g^{-1}\partial_t \hat g)
   \left\{ \hat g^{-1}\tilde\eta g, \hat g^{-1}Q_0g \right\}\right.
\nonumber \\ && \left.~~~~~
  +g^{-1}(Q_0\bar \Psi)g (\tilde\eta\Psi)
  -\frac{1}{3}\bar\Psi(Q_0\bar\Psi)^2
  +\frac{1}{3}\Psi(\tilde\eta\Psi)^2
\right\rangle
\label{S;BISFT}\\
&&\hspace{-3mm}\textrm{with}~~Q_0=-\int \eta\partial\eta e^{2\phi}b,~~
g=e^\Phi,~~\hat g=e^{\hat\Phi(t)},~~\hat\Phi(0)=0,~~\hat\Phi(1)=\Phi,
\nn
\end{eqnarray}
where the operator $Q_0$ is a pure ghost operator
which is related to the BRST operator $Q$ by a similarity transformation \bref{Q}.

As before, $(G+A)^2=0$ implies the equations of motion when
\begin{eqnarray}
  \begin{array}{lcl}
   G=G_0+G_{-1},    && A=A_0+A_{-\frac{1}{3}}+A_{-\frac{2}{3}},   \\
   ~G_0=Q_0,    &&  ~A_0=g^{-1}(Q_0g),  \\
   ~G_{-1}=\tilde\eta,    &&  ~A_{-\frac{1}{3}}=g^{-1}(Q_0\bar\Psi)g,  \\
       &&  ~A_{-\frac{2}{3}}=\tilde\eta\Psi.\\
  \end{array}
\label{PG;A}
\end{eqnarray}
The operator $G$ is nilpotent
because $Q_0^2=0$, $\tilde\eta^2=0$ and $\{Q_0,\tilde\eta\}=0$.
The equation $(G+A)^2=0$ is invariant under the gauge transformation,
$\delta A=G\sigma+[A,\sigma]$, where
\begin{eqnarray}
&&\sigma=\sigma_{\frac{1}{3}}+\sigma_0+\sigma_{-\frac{1}{3}},
\nonumber\\ &&
\sigma_{\frac{1}{3}}=\{Q_0+A_0, \Lambda_{\frac{1}{3}} \},~~
\sigma_0=\tilde\eta\Lambda_1+\{A_{-\frac{2}{3}}, \Lambda_{\frac{2}{3}} \},~~
\sigma_{-\frac{1}{3}}=\tilde\eta\Lambda_{\frac{2}{3}}.
\end{eqnarray}
\medskip

Now, let us examine an expansion around a classical solution
$\Phi_0$, $\Psi_0$ and $\bar\Psi_0$
\begin{eqnarray}
g=g_0h,~~g_0=e^{\Phi_0},~~h=e^\phi,~~\Psi=\Psi_0+\psi,~~\bar\Psi=\bar\Psi_0+\bar\psi.
\label{PG;fluctuation}
\end{eqnarray}
Substituting \bref{PG;fluctuation} into $A$ of \bref{PG;A},
we find that $A$ is expanded as,
$A=\bar A+a$, where
\begin{eqnarray}
  \begin{array}{ll}
\bar A_0=g_0^{-1}(Q_0g_0),
 &~~ a_0=h^{-1}(Qh+[g_0^{-1}(Qg_0),h]),    \\
\bar A_{-\frac{1}{3}}=g_0^{-1}(Q_0\bar\Psi_0)g_0, 
 &~~a_{-\frac{1}{3}}=h^{-1}[g_0^{-1}\bar\Omega_0g_0,h]
  +h^{-1}g_0^{-1}Q\bar \psi g_0h,\\
\bar A_{-\frac{2}{3}}=\tilde\eta\Psi_0,
 &~~a_{-\frac{2}{3}}=\tilde\eta\psi.\\
  \end{array}
  \label{a;original}
\end{eqnarray}
These $a$'s satisfy the equation $(\tilde G+a)^2=0$ with
\begin{eqnarray}
\tilde GX=GX+\{\bar A,X].
\label{Gtilde}
\end{eqnarray}
The nilpotency of the operator $\tilde G$ follows immediately from
the nilpotency $G^2=0$, the equation of motion $G\bar A+\bar A^2=0$
and a Bianchi identity $\{\bar A,\{\bar A,X]]+\{X,\bar A^2]=0$.
The definition \bref{Gtilde} implies that 
we can define operators $\tilde G_n$ by
\begin{eqnarray}
\tilde G_0X&=&G_0X+\{\bar A_0, X],\nonumber\\
\tilde G_{-\frac{1}{3}}X&=&\{\bar A_{-\frac{1}{3}}, X],\nonumber\\
\tilde G_{-\frac{2}{3}}X&=&\{\bar A_{-\frac{2}{3}}, X],\nonumber\\
\tilde G_{-1}X&=&G_{-1}X.
\end{eqnarray}
It is straightforward to see that the nilpotency of $\tilde G$
is expressed as a set of equations satisfied by the operators $\tilde G_n$
order by order of C-charge;
\begin{eqnarray}
&&\tilde G^2_{0}=0,
\nonumber\\&&
\{\tilde G_{0},\tilde G_{-\frac{1}{3}}\}=0,
\nonumber\\&&
\{\tilde G_{0},\tilde G_{-\frac{2}{3}}\}+\tilde G^2_{-\frac{1}{3}}=0,
\nonumber\\&&
\{\tilde G_{0},\tilde G_{-1}\}+\{\tilde G_{-\frac{1}{3}},\tilde G_{-\frac{2}{3}}\}=0,
\\&&
\{\tilde G_{0},\tilde G_{-1}\}+\tilde G^2_{-\frac{2}{3}}=0,
\nonumber\\&&
\{\tilde G_{-\frac{2}{3}},\tilde G_{-1}\}=0,
\nonumber\\&&
\tilde G_{-1}^2=0,\nn
\end{eqnarray}
which imply that the operators $\tilde G_{0}$ and $\tilde G_{-1}$ are
nilpotent.
Using these $\tilde G_n$, we find that $a$'s in \bref{a;original}
can be re-expressed compactly as
\begin{eqnarray}
a_0&=&h^{-1}(\tilde G_0h),\nonumber\\
a_{-\frac{1}{3}}&=&h^{-1}(\tilde G_{-\frac{1}{3}}h)
  +h^{-1}\bar\omega
  h,\label{a}\\
a_{-\frac{2}{3}}&=&\omega,\nn
\end{eqnarray}
where we defined `chiral' and `anti-chiral' string fields by
$\omega =\tilde G_{-1}\psi$ and $\bar\omega=\tilde G_{0}(g_0^{-1}\bar\psi g_0)$.
Note that $a$'s of \bref{a} have the same structure as $A$ of \bref{A} in the section 2.
So we can conclude that the fluctuation string fields
$\phi$, $\omega$ and $\bar\omega$
satisfy the same equations of motion as those
for open RNS super-SFT's of the action \bref{S;SFT2}.
This implies that our super-SFT of the action \bref{S;BISFT}
reproduces  open RNS super-SFT's of the actions \bref{S;SFT2},
if we can find classical solutions, $\Phi_0$, $\Omega_0$ and $\Omegab_0$,
which reproduce the corresponding operator $\tilde G$.
\medskip

In the rest of this section, we concentrate on solving the equation \bref{Gtilde}
in terms of $\Phi_0$, $\Omega_0$ and $\Omegab_0$
with $G$ being 
\begin{eqnarray}
G=Q_0+\etat,
\end{eqnarray}
and $\tilde G$ being $G$ in the equation \bref{G;2}
\begin{eqnarray}
\tilde G&=&e^R(Q+\etat)e^{-R}\nn\\
&=&e^{-U}Q_0e^U
+\cint\eta e^\phi\psi^{-j}\partial x^{+j}
+\cint ce^{-\phi}\psi^{+j}\partial x^{-j}
+\etat,
\end{eqnarray}
where $(\psi^{\pm j},x^{\pm j})=(\psi^{\pm},x^{\pm})$ for d=8
and $j$ runs $1$ to $3$ for d=4.
$U$ is defined in the equation \bref{U}.
\medskip

For this purpose, we introduce the half-string formulation
\cite{W;Non-commutative}
as before,
\begin{eqnarray}
&(A)~~&{\cal I}\star X=X\star{\cal I}=X,~~~\forall X,\label{hs;a}\\
&(B)~~&G^R{\cal I}=-G^L{\cal I},\label{hs;b}\\
&(C)~~&G^R(X)\star Y+(-1)^XX\star G^L(Y)=0,~\forall X,Y,
\label{hs;c}
\end{eqnarray}
where ${\cal I}$ is the identity operator of the algebra.
Using the above definition, one can show \cite{K;Proposal} that
\begin{eqnarray}
\bar A=(\tilde G-G)^L{\cal I}
\label{Abar}
\end{eqnarray}
satisfies the equation (\ref{Gtilde}) because
\begin{eqnarray}
\{\bar A, X]&=&(\tilde G-G)^L{\cal I}\star X-(-1)^XX\star (\tilde G-G)^L{\cal I}
\nonumber\\
&\stackrel{(B)}{=}&-(\tilde G-G)^R{\cal I}\star X-(-1)^XX\star (\tilde G-G)^L{\cal I}
\nonumber\\
&\stackrel{(C)}{=}&{\cal I}\star (\tilde G-G)^L X+(\tilde G-G)^RX\star {\cal I}
\nonumber\\
&\stackrel{(A)}{=}&(\tilde G-G)X
\end{eqnarray}
and the equation $(G+\bar A)^2=0$ because
\begin{eqnarray}
G\bar A&=&(G^L+G^R)(\tilde G-G)^L{\cal I}
=G^L\tilde G^L{\cal I}+G^R\tilde G^L{\cal I}\nonumber\\
&\stackrel{(C)}{=}&G^L\tilde G^L{\cal I}+\tilde G^L\cI\star G^L\cI
\stackrel{(B)}{=}G^L\tilde G^L{\cal I}-\tilde G^R\cI\star G^L\cI\nn\\
&\stackrel{(C)(A)}{=}&G^L\tilde G^L{\cal I}+\tilde G^LG^L\cI
=\{\tilde G^L, G^L\}{\cal I},\\
\bar A^2&=&(\tilde G-G)^L{\cal I}\star (\tilde G-G)^L{\cal I}
\stackrel{(B)}{=}-(\tilde G-G)^R{\cal I}\star (\tilde G-G)^L{\cal I}
\nonumber\\
&\stackrel{(C)(A)}{=}&(\tilde G-G)^L(\tilde G-G)^L{\cal I}
=-\{\tilde G^L, G^L\}{\cal I}\
\end{eqnarray}
where we indicated which equation was used on an equal sign.
In the calculation, we used equations
$(G^L)^2=G^LG^R=(\tilde G^L)^2=0$
which follow from the facts
that the OPE of the two currents $j(z)$ of $G=\oint\hspace{-3.8mm}- j(z)$
is non-singular
\begin{eqnarray}
j(z)j(w)\sim 0,
~~j(z)=-\eta\partial\eta e^{2\phi}b+\eta,
\end{eqnarray}
and that the OPE of the two currents $\tilde j(z)$ of
$\tilde G=\oint\hspace{-3.8mm}- \tilde j$ 
is non-singular
\begin{eqnarray}
\tilde j(z) \tilde j(w)\sim 0,~~
\tilde j=e^{R}(j_{BRST}+\eta)e^{-R}.
\end{eqnarray}
Thus we have shown that the equation \bref{Abar}
solves the equation \bref{Gtilde}.
\medskip

Now
we solve the equation
\bref{Abar} in terms of $\Phi_0$, $\Omega_0$ and $\Omegab_0$.
The definition \bref{a;original} of $\bar A$ implies that the equation \bref{Abar}
can be written, order by order of C-charge, as
\begin{eqnarray}
g_0^{-1}(Q_0g_0)&=&(\tilde G_0^L-Q_0^L){\cal I},\label{eq;1}\\
g_0^{-1}
\bar\Omega_0
g_0&=&\tilde G_{-\frac{1}{3}}^L{\cal I},\label{eq;2}\\
\Omega_0
&=&\tilde G_{-\frac{2}{3}}^L{\cal I},\label{eq;3}\\
0&=&(\tilde G_{-1}-\tilde \eta)^L{\cal I}.
\label{eq;4}
\end{eqnarray}

The first equation \bref{eq;1} is satisfied by
\begin{eqnarray}
\Phi_0=U^L{\cal I},\label{cs1}
\end{eqnarray}
because one can show \cite{K;Proposal} that
\begin{eqnarray}
e^{-U^L{\cal I}}\star Q_0e^{U^L{\cal I}}
&=&\sum_{n=1}^\infty \frac{(-1)^{n-1}}{n!}
[\stackrel{n-1}{\overbrace{U^L{\cal I},...,
[U^L{\cal I} }} ,Q_0(U^L{\cal I})]...] \nonumber\\
&=&\sum_{n=1}^\infty\frac{(-1)^{n-1}}{n!}
[\stackrel{n}{\overbrace{U^L,...,[U^L}},Q_0^L]...]{\cal I} \nonumber\\
&=&(e^{-U}\star Q_0e^{U}-Q_0)^L{\cal I}.
\label{eq;1;cal}
\end{eqnarray}
In the second equality in \bref{eq;1;cal},
we used a relation
\begin{eqnarray}
\{A^R,B^L]=0,~\forall A,B,
\label{[A,B]}
\end{eqnarray}
which follows from the fact that one can show that, using the algebra
\bref{hs;a}-\bref{hs;c},
for any string field $X$,
\begin{eqnarray}
A^RB^LX
&\stackrel{(A)(C)}{=}&-(-1)^{A\cdot BX}B^LX\star A^L\cI
\stackrel{(A)(C)}{=}(-1)^{A\cdot BX}B^R\cI\star X\star A^L\cI\nn\\
&\stackrel{(C)(A)}{=}&-(-1)^{A\cdot BX+A\cdot X}B^R\cI\star A^RX
\stackrel{(C)(A)}{=}(-1)^{A\cdot BX+A\cdot X}B^LA^RX\nn\\
&=&(-1)^{A\cdot B}B^LA^RX.
\end{eqnarray}
In the third equality in \bref{eq;1;cal}, we used a relation,
$\{A,B]^L=\{A^L,B^L]$,
which follows from the equation \bref{[A,B]}.

The second equation \bref{eq;2} is solved by
\begin{eqnarray}
\bar\Omega_0=e^{U^L}\alpha^L e^{-U^L}{\cal I},~~~
\alpha^L=\cint_{L}\eta e^{\phi}\psi^{-j}\partial x^{+j},
\label{cs2}
\end{eqnarray}
because it follows that
\begin{eqnarray}
e^{U^L{\cal I}}\star \alpha^L{\cal I}\star e^{-U^L{\cal I}}
&=&\sum_{n=0}^\infty \frac{1}{n!}
[\stackrel{n}{\overbrace{U^L{\cal I},...,[U^L{\cal I}}},\alpha^L{\cal I}]...] 
\nonumber\\
&=&\sum_{n=0}^\infty \frac{1}{n!}
[\stackrel{n}{\overbrace{U^L,...,[U^L}},\alpha^L]...]{\cal I}
\nonumber\\  
&=&e^{U^L}\alpha^L e^{-U^L}{\cal I}.
\end{eqnarray}

The third equation \bref{eq;3} implies that
\begin{eqnarray}
\Omega_0=\beta^L{\cal I},~~~
\beta^L=\cint_{L}ce^{-\phi}\psi^{+j}\partial x^{-j}.
\label{cs;3}
\end{eqnarray}

The last equation \bref{eq;4}
is automatically satisfied because $\tilde G_{-1}=\tilde \eta$.
\medskip

As a result, manifest d=8,4 Lorentz covariant open RNS super-SFT's
are shown to be obtained
from the pregeometrical super-SFT (\ref{S;BISFT})
using a set of classical solutions (\ref{cs1}), (\ref{cs2}) and (\ref{cs;3}).
The uniqueness of the classical solution for the particular $\tilde G$
can be shown as in \cite{HLRS;Purely}.

\sect{Summary}
In this paper, we proposed a pregeometrical formulation
for Berkovits' manifest d=8,4 Lorentz covariant open RNS super-SFT's.
The pregeometrical action
contains a pure ghost operator $Q_0$ instead of the BRST operator $Q$.
We showed that manifest d=8,4 Lorentz covariant open RNS super-SFT's
are derived from the pregeometrical super-SFT
by expanding around solutions of the equations of motion.
The pregeometrical action takes a different form from
the manifest d=8,4 Lorentz covariant open RNS super-SFT actions
as in the pure cubic SFT \cite{HLRS;Purely},
while  the Berkovits' NS super-SFT action and the pregeometrical action
had the same form \cite{K;Proposal}.
\medskip

We derived open RNS super-SFT's from the pregeometrical super-SFT
at the level of their equations of motion.
We expect that this can be done
at the level of actions also.
\medskip

We used the operator $G$ of \bref{G;2} which leads to
manifest d=4 Lorentz covariant RNS super-SFT,
while Berkovits showed that N=1 d=4 super-Poincare covariant
RNS super-SFT can be obtained using a similarity transform
$\hat G=e^{\frac{1}{2}U}Ge^{-\frac{1}{2}U}$ in \cite{B;Ramond}.
It is interesting to examine whether our $G$
can be rewritten in an N=1 d=4 super-Poincare covariant notation
using Green-Schwarz-like variables
and, if this can be done, whether the obtained N=1 super-Poincare covariant action
can be related to Berkovits' N=1 super-Poincare covariant action.

\medskip

Our formulation is background independent only formally,
because the star product was treated as an abstract
object here,
but the background dependence enters in the concrete expression
of the star product.

\medskip

The C-charge assignment on string fields causes the world-sheet `matter'
superconformal fields to carry C-charge.
The background dependence of the BRST operator $Q$
was eliminated by using the pure ghost operator $Q_0$.
Though the concrete form of the string fields is determined
after the choice of the classical solution,
the C-charge assignment will impose some restriction
on the superconformal `matter' fields dependence of the string fields.
In this way, some background dependence may enter.
This will be related to the fact that
not all N=1 c=15 superconformal field theory backgrounds admit to construct
an open RNS super-SFT action as was mentioned in \cite{B;Ramond}.

\medskip

In this paper, we considered GSO(+) projected string fields only.
So the theory contains BPS objects only.
In order to be truly background independent, the pregeometrical theory
must contain non-BPS objects as well as BPS objects.
This implies that
we need to extend the theory to include GSO(--) string fields.
Such an extension for Berkovits' open \textit{NS} super-SFT was given
and shown to be very useful in the calculation of the tachyon potential
in \cite{B;Tachyon}.
When Berkovits' open \textit{RNS} super-SFT's are extended 
to include GSO(--) projected string states,
this theory will allow us to gain
new insights into tachyon condensation,
including the change of the number of the space-time supersymmetry generators.
We believe that the corresponding pregeometrical formulation for this
can be found along the line presented in this paper.
Such a theory will be more fundamental background independent
formulation,
which describes non-BPS objects as well as BPS objects.
\medskip

We used the operator $Q_0$ which has the non-trivial cohomology,
so that the theory describes the physical open string excitations.
Alternatively, if we choose a ghost $c$ as $Q_0$,
then the theory is on the closed string
vacuum, which is an end point of a tachyon condensation,
because $c$ has the trivial cohomology
and so the theory does not describe any open string excitations.
Recently, such a theory has attracted much attention
as the vacuum SFT's \cite{vacuum}
which can be viewed as a pregeometrical formulation
of the Witten's bosonic SFT.
We expect
that our action with $Q_0=c$ may provide
an action for a vacuum open RNS super-SFT.
It is interesting to examine this theory.
\medskip

Covariant closed super-SFT without picture changing operators
is not known at this moment.
But some suggestions were given using the N=4 topological prescription
\cite{B;New Description,B;Ten-Dimensional}
in \cite{B;New}
and respecting the C-charge assignment in \cite{B;Ramond}.
The pregeometrical formulation of closed super-SFT
will provide more fundamental theory of them.
It is interesting to formulate this theory
in order to gain some
insights into the
structures of closed super-SFT's.

\vspace{6mm}

\begin{flushleft}
{\large \textbf{Acknowledgment}}\\
The author would like to express his gratitude to
the theory group of KEK.
\end{flushleft}



\end{document}